\documentclass[apj]{emulateapj}
\usepackage{apjfonts}

\usepackage{amsmath}

\shorttitle{PULSATIONS OF HS 1824+6000}
\shortauthors{STEINFADT ET AL.}

\newcommand{\be}{\begin{eqnarray}}
\newcommand{\ee}{\end{eqnarray}}

\begin{document}

% -----------------------------------------------------------
% -----------------------------------------------------------

\title{Pulsations of the Low Mass ZZ Ceti Star HS 1824+6000}

\author{Justin D. R. Steinfadt}
\affil{Department of Physics, Broida Hall,\\University of California,
Santa Barbara, CA 93106;\\jdrs@physics.ucsb.edu}

\author{Lars Bildsten}
\affil{Kavli Institute for Theoretical Physics and Department of Physics, Kohn Hall,\\University of California, Santa Barbara, CA 93106;\\bildsten@kitp.ucsb.edu}

\author{Eran O. Ofek, Shri R. Kulkarni}
\affil{Division of Physics, Mathematics and Astronomy,\\California Institute of Technology, Pasadena, CA 91125}

% -----------------------------------------------------------
% -----------------------------------------------------------

\begin{abstract}

Measuring g-mode pulsations of isolated white dwarfs can reveal their interior properties to high precision. With a spectroscopic mass of $\approx 0.51 M_{\odot}$ ($\log g = 7.82$), the DAV white dwarf HS 1824+6000 is near the transition between carbon/oxygen core and helium core white dwarfs, motivating our photometric search for additional pulsations from the Palomar 60-inch telescope.  We confirmed (with much greater precision) the three frequencies: $2.751190 \pm 0.000010$ mHz (363.479 sec), $3.116709 \pm 0.000006$ mHz (320.851 sec), $3.495113 \pm 0.000009$ mHz (286.114 sec), previously found by B. Voss and collaborators, and found an additional pulsation at $4.443120 \pm 0.000012$ mHz (225.067 sec). These observed frequencies are similar to those found in other ZZ Ceti white dwarfs of comparable mass (e.g. $\log g<8$).  We hope that future observations of much lower mass ZZ Ceti stars ($< 0.4 M_{\odot}$) will reveal pulsational differences attributable to a hydrogen covered helium core.
\end{abstract}


% -----------------------------------------------------------
% -----------------------------------------------------------

\section{Introduction}

% TABLE ONE
\begin{deluxetable}{lcc}
\tablewidth{0pt}
\tablecaption{Properties of HS 1824+6000\label{tab:prop}}
\tablehead{
\colhead{ } & \colhead{\citealt{bv06}} & \colhead{\citealt{ag07}} }
\tablecolumns{3}
%\scriptsize
\startdata
	$T_{\rm eff}$ (K) & $11192\pm300$\tablenotemark{a} & $11380\pm140$\tablenotemark{b} \\
	$\log g$ (dex) & $7.65\pm0.10$\tablenotemark{a} & $7.82\pm0.04$\tablenotemark{b} \\
	$m_B$ (mag) & 15.7 & 15.7 \\
	\cutinhead{Observed Frequencies (in mHz)}
	\citealt{bv06} & \multicolumn{2}{l}{$2.6\pm0.4$, $3.0\pm0.9$, $3.3\pm0.4$, $3.4\pm0.8$} \\
	This Paper & \multicolumn{2}{l}{$2.751190 \pm 0.000010$, $3.116709 \pm 0.000006$} \\
	& \multicolumn{2}{l}{$3.495113 \pm 0.000009$, $4.443120 \pm 0.000012$}
\enddata
\tablenotetext{a}{Photometrically determined.}
\tablenotetext{b}{Spectroscopically determined.}
\end{deluxetable}

% FIGURE ONE
\begin{figure*}[t]
	\centering
	\epsscale{1.0}
	\plotone{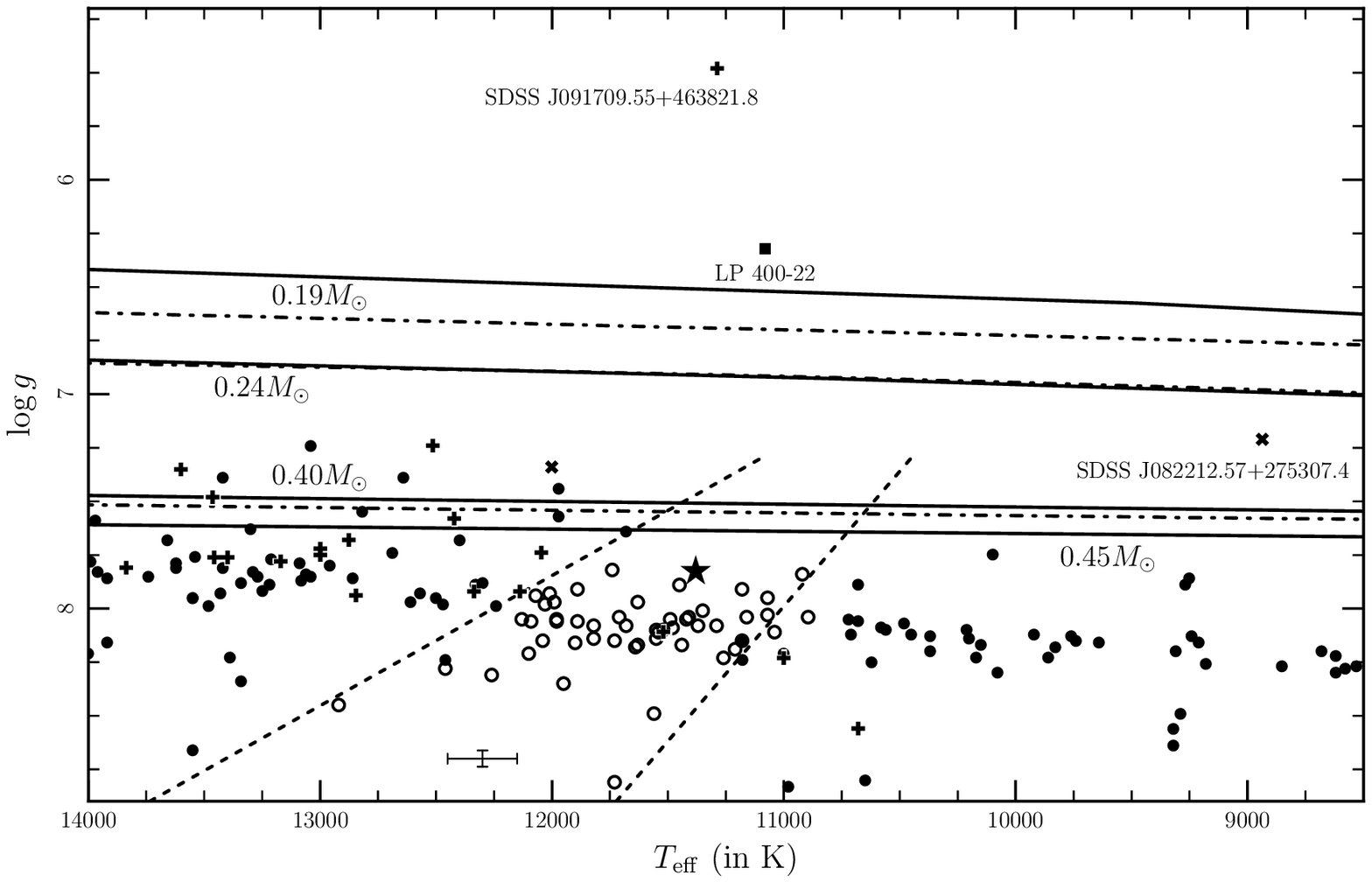}
	\caption{The empirical ZZ Ceti instability strip.  The circles represent systems for which temporal observations have been performed.  Filled circles indicate systems not observed to vary while open circles indicate systems with observed periods.  These data are from \citet{pb04} and \citet{ag05,ag07}.  The vertical crosses represent low-mass WDs selected from the SDSS with $\log g$ and $T_{\rm eff}$ redetermined from MMT spectra from \citet{mk07a}.  The diagonal crosses are from \citet{mk07a} except they are reanalysis of SDSS spectra and are merely candidate low-mass WDs until better spectra can be obtained.  The square is LP 400-22 \citep{ak06,mk07a}.  The star is HS 1824+6000 (see Table \ref{tab:prop}, \citealt{ag07}).  The error bar in the instability strip represents the typical error for measurements within the instability strip.  The dashed lines are empirical fits to the instability strip as determined by \citet{ag07}.  The solid \citep{jap07} and dash-dotted \citep{lga01} lines correspond to cooling tracks of He WDs of the labeled mass.  The labeled masses correspond to the following model masses (in $M_{\odot}$) for \citet{jap07} and \citet{lga01} respectively: 0.19 to 0.1869 and 0.196, 0.24 to 0.2495 and 0.242, 0.40 to 0.3986 and 0.406, 0.45 to 0.4481 (\citealt{jap07} only).} % 0.16 to 0.1604 and 0.161,
	\label{fig:instrip}
\end{figure*}

The hydrogen line ZZ Ceti variable (DAV) white dwarfs (WDs) occupy a discrete strip in the $T_{\rm eff}-\log g$ plane known as the ZZ Ceti instability strip.  Many groups have assessed the location of this instability strip both empirically \citep{fw91,asm04b,ag05,bgc07} and theoretically \citep{pb97,yw99,gf03}, and despite minor discrepancies it spans $11,000 \lesssim T_{\rm eff} \lesssim 12,250$ K for $\log g \approx 8.0$.  \citet{gf82} suggested that the instability strip is pure, meaning that all WDs within the strip are variable.  However, \citet{asm05} found numerous objects from the SDSS with associated low signal-to-noise spectra that were tenuously identified to be in the strip but did not vary to their observed detection limits.  Recent observations \citep{bgc07} have found some of these to be low amplitude pulsators.  If the instability strip is pure, then it strongly implies that ZZ Ceti stars are a phase of evolution through which all DA WDs must evolve as they cool.

White dwarfs less massive than $\approx 0.45-0.47 M_{\odot}$ ($\log g \approx 7.67$ at $T_{\rm eff} \approx 11,500$ K) \citep{nld96,id99,ap04,dav06,jap07} do not undergo a He core flash in the course of their evolution and therefore are left with a He core.  Two modes of evolution can truncate the red giant branch evolution and prevent the He core flash:  mass loss due to winds and mass loss due to binary interaction.  In systems of high metallicity, mass loss due to stellar winds on the red giant branch can be significant enough to lose the H envelope prior to the core flash \citep{nld96,bmh05,mk07b}.  Binary interaction through a common envelope also leads to significant mass loss \citep{ii93,trm95}.  Helium is thus the expected core composition for WDs below $\approx 0.45-0.47 M_{\odot}$.  However, little direct evidence exists of the He core.  Possible evidence would be the apparent over-brightness of old WDs \citep{bmh05} in the star cluster NGC 6791 \citep{lrb05}.  Though uncertainties remain \citep{cjd02,lrb08a,lrb08b}, recent detection of low $\log g$ young WDs \citep{jsk07} makes it plausible for many of the old WDs to be He core.  A detailed asteroseismological study of these low-mass WDs could provide convincing evidence for the core composition.  Observation and analysis of a full spectrum of the pulsation modes in a WD can produce a wealth of information about the interior structure of the WD.  The mean period spacing of the modes, the rate of change in a mode's period over time, and multiplet splitting of individual modes can provide information on the total mass, spin rate, magnetic field strength, mass of H envelope, and core composition of the WD \citep{ahc02,bgc08}.  This has already been theoretically applied to distinguish between C/O and O/Ne core WDs by \citet{ahc04}.  The measured change in an observed mode period in G117-B15 has also been used to constrain significantly the C/O core composition of this object \citep{sok91,sok95,sok00,sok05a}.

We plot a version of the empirical instability strip in Figure \ref{fig:instrip} .  Included are not observed to vary (NOV) systems and pulsating ZZ Ceti stars from the observations of \citet{pb04} and \citet{ag05,ag07}.  Also included are low-mass WDs from \citet{mk07a}.  There is a notable absence of low-mass ($\log g \lesssim 7.67$) WDs within the instability strip.  There are a few possible ZZ Ceti stars of this low mass that are not plotted due to the absence of spectroscopically determined $\log g$ and $T_{\rm eff}$ measurements \citep{bv07}.  Also shown are the He WD cooling tracks of two models for $\approx 0.19$, 0.24, 0.40, and $0.45 M_{\odot}$ WDs \citep{lga01,jap07}.  The difference between these two models at low mass is due to the different H envelope masses.  \citet{lga01} used a $1 M_{\odot}$ main sequence star and truncated its evolution up the red giant branch at various stages to produce He WDs of varying masses.  \citet{jap07} used close binary evolution expectations for main sequence stars of many masses to produce He WDs of varying masses.  These different approaches cause the different remnant H envelope masses that yield a degeneracy in the He WD mass and its position in the $T_{\rm eff}-\log g$ plane.  This degeneracy would be broken in the case of a ZZ Ceti He WD where the pulsation mode spectrum would reveal the H envelope mass. 

HS 1824+6000 (hereafter HS 1824, see Table \ref{tab:prop}) was initially observed by \citet{bv06} to exhibit pulsations.  Their photometrically determined $\log g$ and $T_{\rm eff}$ placed its mass at $\approx 0.40 M_{\odot}$ using the tables of \citet{lga97}.  This mass was well within the theoretical expected mass range for He core WDs making it an excellent object to compare and contrast its pulsation frequencies with other C/O core and possible He core DAVs.  However, later spectroscopic measurement by \citet{ag07} determined its mass to be $\approx 0.51 M_{\odot}$, beyond the expected mass range for He core WDs.  In \S\ref{sec:obsana} we discuss our own observations and differential photometry of HS 1824.  In \S\ref{sec:timing} we apply a Lomb-Scargle Periodogram approach to a non-uniformly sampled time series in order to obtain the pulsation frequencies of HS 1824.  In \S\ref{sec:flcpgs} we report the results of our observations and analysis.

In \S\ref{sec:conc} we compare all observed ZZ Ceti periods with $\log g < 8.0$.  It is our hope this will yield a `zeroth-order' approach to He core identification in much the same way $T_{\rm eff}$ and $\log g$ measurements of field WDs identify likely ZZ Ceti stars.  No singular distinction is currently present.

% -----------------------------------------------------------
% -----------------------------------------------------------

\clearpage
\section{Observations}
\label{sec:obsana}

We observed HS 1824 (see Table \ref{tab:prop} for properties) on eleven nights from 2006 August to 2006 October using the robotically operated 60-inch (1.52 m) telescope at the Palomar Observatory \citep{sbc06}.  All observations had 30 second exposures with dead times ranging from 20-40 seconds.  To reduce dead time, half the CCD was read out.  The large variance in the dead time was due primarily to a technical problem within the automated observing software used to control the telescope.  The observing durations varied from 1 - 3 hours.  The primary 2048 $\times$ 2048 pixel, 11' $\times$ 11', CCD for the robotic Palomar 60-inch was used in all observations with a Gunn $g$ filter.  We chose the Gunn $g$ filter to optimize the ratio of pulsation count amplitude to total stellar counts.  It is a known trend that this ratio is larger in bluer filters such as Gunn $g$ \citep{elr82,elr95}.  A clear filter would not be optimal as it increases the total stellar counts without a comparable increase in pulsation count amplitude, thus reducing this important ratio.  Flat fielding, bias subtraction, and sky subtraction were performed within the data pipeline of the Palomar 60-inch Telescope Archive \citep{sbc06}.  The sky subtraction was done as an inaccurate scalar value and for our purposes was added back into the data and recalculated using standard, more accurate IRAF\footnote{IRAF (Image Reduction and Analysis Facility) is distributed by the National Optical Astronomy Observatory, which is operated by the Association of Universities for Research in Astronomy, Inc., under contract with the National Science Foundation.  http://iraf.noao.edu} tools.

% -----------------------------------------------------------
% -----------------------------------------------------------

\subsection{Data Reduction}
\label{sec:datared}

We used the IRAF package VAPHOT \citep{hjd01} to dynamically determine optimum aperture sizes as a function of seeing for our photometry.  Given a characteristic frame for each night, VAPHOT calculates an optimized aperture using a PSF for each star that maximizes the signal to noise within the aperture.  This optimized aperture is then found as a function of the seeing.  Then, for a time series of frames, VAPHOT calculates the seeing value for each individual frame and scales the optimized apertures accordingly.  Finally, all aperture information is input into the standard IRAF task \textit{phot} which calculates counts within the optimized aperture along with background noise counts measured in an annulus just beyond the optimized aperture.  For each night of observation, 21 comparison stars were selected ranging $g=12-16$ mag and along with the program star, counts and background information were extracted using the VAPHOT task.  Additionally, the exposure start times for all frames of observation were converted to barycentric Julian dates.

The uncertainty of this photometry for each aperture is given by;
\be
	\sigma_{CCD}^2=c+n_{bins}\left(1+\frac{n_{bins}}{n_{sky}}\right)(N_S+N_R^2+N_D),
\ee
\noindent \citep{sbh06}, where $c$ is the number of integrated source counts in photons, $n_{bins}$ is the aperture area calculated by VAPHOT in pixels, $n_{sky}$ is the area of the annulus used to calculate the background information in pixels, $N_S$ is the background counts per pixel, $N_R$ is the read noise of the CCD in counts per pixel, and $N_D$ is the dark current in counts per pixel.  We do not include the digitization error as it is significantly less than our value for the gain.  For our observations, $c$ was $\approx 4\times10^4$ counts for HS 1824 and $\approx 10^4 - 10^6$ counts for the comparison stars, while $N_S \approx 40-600$ counts, $N_R^2 \approx 25-60$, and $N_D \ll 1$ counts for  $t_{int}=30$ sec integrations.

% TABLE TWO
\begin{deluxetable*}{cccccccc}[t]
\tablewidth{0pt}
\tablecaption{Observation Dates and Pulsation Results for HS 1824+6000\label{tab:freq}}
\tablehead{
\colhead{Date (UT)} & \colhead{2.7 (Ampl.)} & \colhead{3.1 (Ampl.)} & \colhead{3.5 (Ampl.)} & \colhead{4.4 (Ampl.)} & \colhead{Obs. Length\tablenotemark{a}} & \colhead{Number of} & \colhead{Number of} \\
\colhead{YYMMDD} & \colhead{mHz (mmag)} & \colhead{mHz (mmag)} &
\colhead{mHz (mmag)} & \colhead{mHz (mmag)} & \colhead{hr} & \colhead{Frames\tablenotemark{a}} & 
\colhead{Comp. Stars\tablenotemark{b}} }
\tablecolumns{7}
\scriptsize
\startdata
	060821 & - & - & - & $4.45 \pm 0.10$ (9.7) & 0.8 & 56 & 12 \\
	060824 & $2.68 \pm 0.16$ (8.1) & $3.15 \pm 0.16$ (9.1) & - & - & 0.9 & 59 & 13 \\
	060827 & - & $3.08 \pm 0.15$ (10.4) & - & - & 0.9 & 59 & 14 \\
	060830 & - & $3.14 \pm 0.13$ (6.2) & - & $4.44 \pm 0.13$ (5.9) & 1.1 & 69 & 11 \\
	060903 & - & - & - & - & 1.0 & 68 & 11 \\
	060906 & - & $3.13 \pm 0.14$ (7.8) & - & - & 1.0 & 69 & 14 \\
	060909 & - & $3.15 \pm 0.12$ (11.8) & - & - & 1.2 & 66 & 10 \\
	061009 & $2.74 \pm 0.05$ (5.6) & $3.13 \pm 0.05$ (7.5) & - & $4.45 \pm 0.05$ (7.0) & 2.8 & 137 & 9 \\
	061016 & - & $3.10 \pm 0.03$ (8.1) & $3.53 \pm 0.03$ (5.3) & - & 3.5 & 129 & 9 \\
	061019 & $2.73 \pm 0.03$ (7.4) & $3.14 \pm 0.03$ (7.2) & - & - & 4.0 & 164 & 7 \\
	061021 & - & $3.10 \pm 0.09$ (8.9) & $3.55 \pm 0.09$ (5.6) & - & 1.5 & 90 & 11 \\
	\cutinhead{Weighted Average over Four Longest Nights}
	\nodata & $2.73 \pm 0.03$ & $3.12 \pm 0.02$ & $3.53 \pm 0.03$ & $4.45 \pm 0.05$ & \nodata & \nodata & \nodata \\
	\cutinhead{Combined Data Set (See \S\ref{sec:flcpgs})}
	\nodata & \multicolumn{1}{l}{$2.751190$} & \multicolumn{1}{l}{$3.116709$} & \multicolumn{1}{l}{$3.495113$} & \multicolumn{1}{l}{$4.443120$} & \nodata & \nodata & \nodata \\
	& \multicolumn{1}{r}{$\pm 0.000010$} & \multicolumn{1}{r}{$\pm 0.000006$} & \multicolumn{1}{r}{$\pm 0.000009$} & \multicolumn{1}{r}{$\pm 0.000012$} & & & \\
\enddata
\tablenotetext{a}{Includes only data used in analysis; excludes contaminated frames (i.e. cosmic ray in program star, clouds, poor seeing, etc.).}
\tablenotetext{b}{See \S\ref{sec:flcpgs} for discussion on the inclusion and exclusion criteria for comparison stars. There was a maximum 21 comparison stars possible.}
\tablecomments{`-' : denotes the frequency was not observed to the required significance level of 90\%}
\end{deluxetable*}

Additional uncertainty arises from atmospheric variability on spatial scales of the CCD field of view.  Scintillation is a dimensionless measure of the flux variations of a source observed through a finite aperture (our telescope) due to fluctuations in the refractive index of the atmosphere caused by temperature changes.  Young's formulation \citep{aty67} of Reiger's theory of scintillation \citep{shr63} gives $s_{scint}=S_0 d^{-2/3} X^{3/2} e^{-h/h_0} \Delta f^{1/2}$, where $S_0=0.09$ is a constant \citep{aty67}, $d=152$ cm is the mirror diameter, $X$ is the airmass, $h=1706$ m is the Palomar Observatory altitude, $h_0=8000$ m is a constant \citep{aty67}, and $\Delta f =1/t_{int}$.  The formal photometric error for each star in each frame is,
\be
	\sigma^2=\sigma_{CCD}^2+s_{scint}^2 c^2, \label{eqn:err}
\ee
\noindent which determines the count level, $c$, at which scintillation noise becomes comparable to Poisson noise.  This occurs at $9\times10^5$ counts at airmass 1.15 and $2\times10^5$ counts at airmass 2.0.  Compared to the total formal error given by Equation (\ref{eqn:err}), scintillation accounts for $10-40\%$ of the error depending upon the airmass (higher airmass account for higher percentages).  Therefore, we are mostly limited by Poisson counting statistics, but scintillation can become significant at higher airmass.

% -----------------------------------------------------------
% -----------------------------------------------------------

\subsection{Differential Photometry}
\label{sec:diffphot}

Since the atmosphere is constantly changing, differential rather than absolute photometry was used in the construction of our light curves.  We used an ensemble of comparison stars to reduce the noise level inherent in any single comparison star.  We used the weighting scheme detailed in \cite{jls01} inspired by \cite{rlg88}.  For our target star we define;
\begin{mathletters}
\begin{eqnarray}
	x(i) &=& A \frac{c_p(i)}{\sum_{m=1}^K w_m c_m(i)}, i=1,...,N, \\
	\sigma_x^2(i) &\approx& \left[\frac{\sigma_p(i)}{c_p(i)} \right] + \frac{\sum_{m=1}^{K} [w_m \sigma_m(i)]^2}{\left[ \sum_{m=1}^{K} w_m c_m(i) \right]^2},\\
A^{-1} &=& \frac{1}{N} \displaystyle\sum_{i=1}^{N} \frac{c_p(i)}{\sum_{n=1}^{K} w_n c_n(i)}, \\
	w_m &=& \frac{\sum_{i=1}^{N} c_m(i)}{\sum_{i=1}^{N} \sigma_m^2(i)},
	\end{eqnarray}
\end{mathletters}

\noindent where $x(i)$ is the count ratio for the $i$'th image, $c_p(i)$ and $\sigma_p(i)$ are the background-subtracted counts and uncertainty of the program star, $c_m(i)$ and $\sigma_m(i)$ are the background-subtracted counts and uncertainty for the $m$'th comparison star in the $i$'th image, $K$ is the number of comparison stars, and $N$ is the total number of frames in the light curve.  The weights of the $m$'th comparison star, $w_m$, are the same for every image, while $A$ is a normalization factor that gives $x(i)$ meaning such that;
\be
	\Delta c_p(i)=\bar{c_p} \left( x(i)-1 \right),
\ee
\noindent where $\bar{c_p}$ is the mean background subtracted counts of the program star and $\Delta c_p(i)$ is the difference in the total counts of the $i$th frame compared to the mean counts of the program star for that night.

% -----------------------------------------------------------
% -----------------------------------------------------------

\section{Lomb-Scargle Timing Analysis}
\label{sec:timing}

The robotically controlled Palomar 60-inch presents a few challenges for time domain observations.  First is the variable dead time of 20-40 seconds after a 30 second exposure.  Second, the automated observing program sometimes places a higher priority on other targets, thus placing temporal gaps in our time series.  While data gaps can be addressed in discrete Fourier analysis, large variations in timing is a much more difficult problem that we address via the Lomb-Scargle periodogram approach.

\cite{jds82} defines a periodogram as a function of the angular frequency $\omega$ (in $\rm rad \; s^{-1}$) as follows;
\begin{mathletters}
\begin{eqnarray}
	P_x(\omega) &=& \frac{1}{2} \left( \frac{\left[ \sum_{i=1}^N x(t_i) \cos(\omega(t_i-\tau)) \right]^2}{\sum_{i=1}^N \cos^2(\omega(t_i-\tau))} \right. \nonumber \\
	&& \left. + \frac{\left[ \sum_{i=1}^N x(t_i) \sin(\omega(t_i-\tau)) \right]^2}{\sum_{i=1}^N \sin^2(\omega(t_i-\tau))} \right), \\
	\tan(2 \omega \tau) &=& \frac{\sum_{i=1}^N \sin(2 \omega t_i)}{\sum_{i=1}^N \cos(2 \omega t_i), }
\end{eqnarray}
\end{mathletters}

\noindent where $x(t_i)$ is $\Delta c_p(i)$ from our differential photometry (\S\ref{sec:diffphot}) and $t_i$ is the start time in seconds of the $i$'th frame.  Further considerations by \cite{jds82} and \cite{jhh86} showed that the probability distribution of power at frequency $\omega$ (with Gaussian white noise) is $\mathrm{Prob}(\mathrm{P}(\omega)>z)=e^{-z}$ when the periodogram is normalized as
\be
	P(\omega)=P_x(\omega)/\sigma^2,
\ee
\noindent where $\sigma^2$ is the total measured variance of $x(t)$ over the entire time series. \cite{jhh86} also showed that for data with periodic signals, the normalization factor remains the total variance of the raw data with the signal present.

The $\exp (-z)$ probability distribution quantifies the significance of any signal seen within the periodogram, allowing us to find the probability that the noise (presumed to be independent and normal) would, by itself, produce a power of $z$.  This allows us to generate a \textit{false-alarm probability} that states if we scan $M$ independent frequencies then the probability that the intrinsic noise produces a power greater than $z$ in any one of the frequency bins is
\be
	\mathrm{Prob}(\mathrm{Any \, Power} \, > \, z)=1-(1-e^{-z})^M.
	\label{eq:falarm}
\ee
\noindent A periodic signal is thus significant to 90\% over all $M$ sampled frequencies if the false-alarm probability is 10\%.  Since our noise is not exactly normally distributed due to the presence of unresolved pulsations which assure some correlation between data point the precise significance may be slightly lower than this.  However, this will not have any consequence for our results.

% -----------------------------------------------------------
% -----------------------------------------------------------

% FIGURE TWO
\begin{figure*}
	\centering
	\epsscale{1.0}
	\plotone{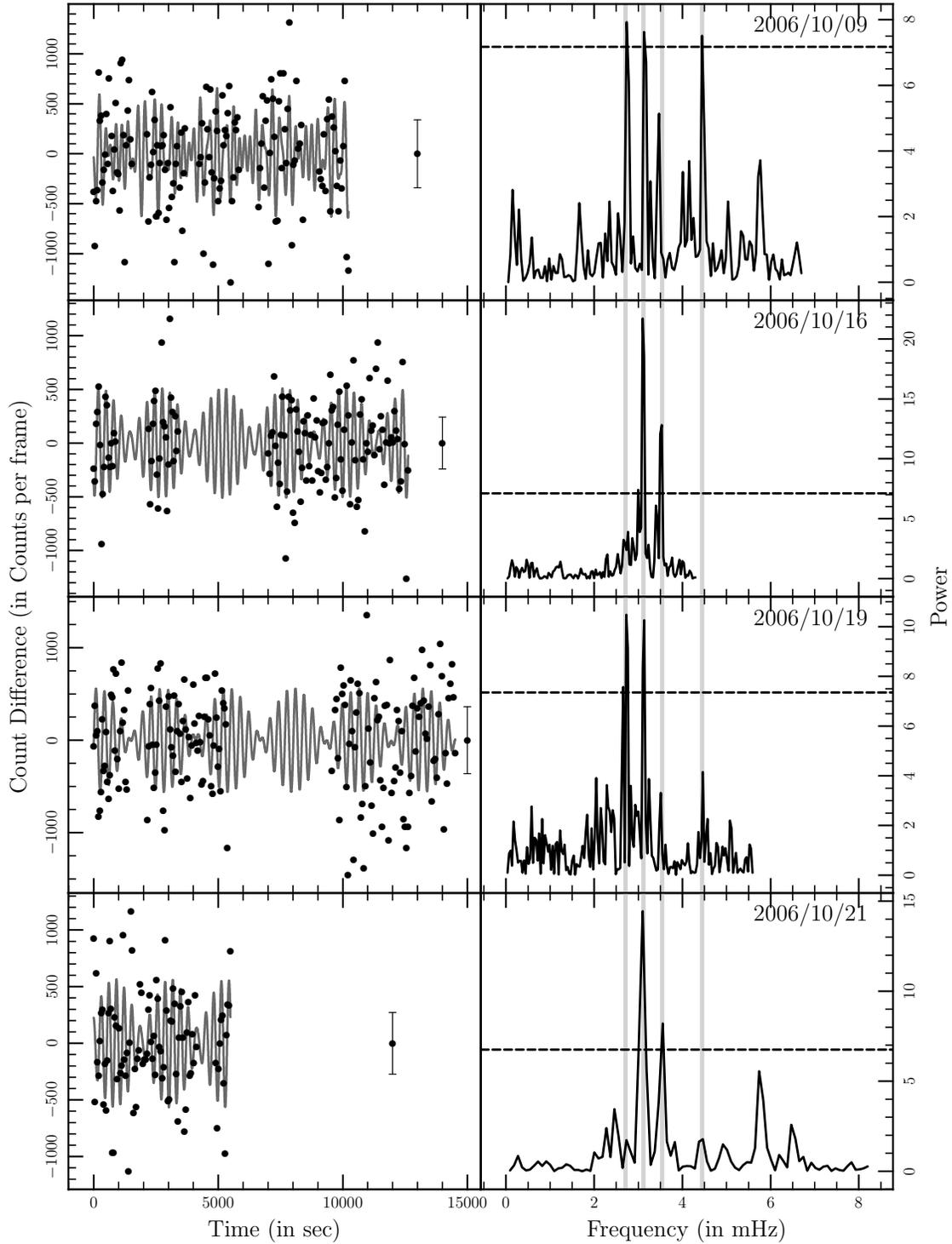}
	\caption{Left Panels: Light curves (see \S\ref{sec:diffphot}) plotted as count difference compared to mean counts of the program star on that night versus time.  Counts refers to the total counts accumulated within the 30 second exposure.  A least-squares fit of sinusoidal functions to the observed significant frequencies is also plotted.  Points detached from data sets denote typical error bars.  Right Panels: Lomb-Scargle Periodograms. (see \S\ref{sec:timing})  Dashed lines denote power level required for 90\% significance.  Gray vertical lines show locations of 2.7, 3.1, 3.5, and 4.4 mHz frequencies.}
	\label{fig:lcpsd}
\end{figure*}

\section{Final Light Curves and Periodograms: Results}
\label{sec:flcpgs}

All 21 comparison stars would not produce the most stable comparison set.  To determine the optimum ensemble of comparison stars for a given observing run, every comparison star was compared to all other comparison stars one at a time by calculating light curves (\S\ref{sec:diffphot}) and periodograms (\S\ref{sec:timing}).  Such an analysis reveals consistent frames where a comparison star has a count value much beyond the scatter of the normal light curve.  In these cases, that frame and comparison star were analyzed using standard IRAF tasks to determine what caused the contamination (e.g. cosmic ray strike, drift into bad pixel due to poor guiding).  Almost always, these comparison stars were then excluded from the optimum comparison star ensemble for that night only.  The periodograms found those comparison stars with consistent frequency content due to possible intrinsic variability.  These comparison stars were also excluded.  Individual frames were excluded when the program star was contaminated by cosmic ray strikes, or the entire frame was affected by a high background level or poor seeing.

Comparison stars were also excluded when color-airmass effects could not be adequately removed via a de-trending second order polynomial.  This was done by comparing all comparison stars to HS 1824 individually and looking for high levels of noise in the lowest frequency domain of the periodogram.  De-trending with a second order polynomial is acceptable in our situation as the periods of pulsation are much shorter than the hours time-scale it takes for changing color-airmass. The resulting optimum comparison star ensemble was then used to compute the HS 1824 differential light curve (\S\ref{sec:diffphot}), which was de-trended through the second order polynomial fitting, and the final light curve processed through the periodogram (\S\ref{sec:timing}).

Using Equation (\ref{eq:falarm}), we determined an observed power to be significant in any periodogram if the probability was greater than 90\% (\textit{false-alarm} probability less than 10\%) over all sampled frequencies.  A summary of all significant frequencies is in Table \ref{tab:freq}, where the frequency uncertainty reported is the separation of the frequency bins in the periodogram.  Figure \ref{fig:lcpsd} shows the differential light curves and periodograms for our four longest data sets.  The data from the four longest observations allow us to construct a weighted average to arrive at $2.73\pm0.03$ mHz (366 sec), $3.12\pm0.02$ mHz (321 sec), $3.53\pm0.03$ mHz (283 sec), and $4.45\pm0.05$ mHz (225 sec).  The 2.7 and 4.4 mHz frequencies are confirmed in three nights, the 3.5 mHz frequency in two nights, and the 3.1 mHz frequency in nine nights.  Excess power is often observed in these frequencies on other nights although not to the required significance level (90\%).

\begin{figure}
	\centering
	\epsscale{1.0}
	\plotone{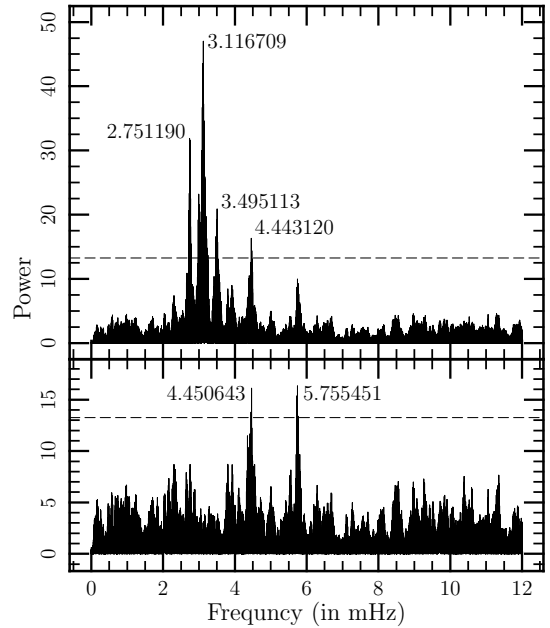}
	\caption{Top Panel: Lomb-Scargle Periodogram for combined data set (see \S\ref{sec:flcpgs}).  Bottom Panel: Lomb-Scargle Periodogram for de-signaled (using only the four significant to 90\% detected frequencies) data set .  For both plots the dashed lines denote power level required for 90\% significance.}
	\label{fig:cspec}
\end{figure}

Our final analysis combined all eleven nights of data into one data set.  Sky conditions were not the same for all nights, so the individually reduced data as described above was used and then combined.  Barycentric Julian dates must be used in this analysis as changes in the Earth's orbital position in the solar system can account for as much as a eight seconds per day change in light arrival time. This composite data set was spectrally analyzed using the Lomb-Scargle periodogram and the result is plotted in Figure \ref{fig:cspec}.

All four detected frequencies are recovered to our 90\% confidence, however, the imprint of our window function makes it difficult to determine any gains in precision over the individual nights.  To address this concern we used a method of least-squares fitting of sinusoids at all four detected frequencies, allowing a single frequency to vary while fixing the remaining frequencies and minimizing $\chi^2_{fit}$.  The old detected frequency was then replaced with this more accurate frequency.  This was done for all four frequencies and repeated recursively until all four frequencies no longer changed values significantly.  This method gives us accurate determinations of the frequencies, amplitudes, and phases of the four detected frequencies.  To determine the precision of these new measurements we used a more robust $\chi^2_{fit}$ minimization technique allowing all parameters to vary, now including the frequencies.  The inherent non-linearity of the fitting model requires the use of the Levenberg-Marquardt method which is given the accurate determinations of the frequencies, amplitudes, and phases of the four detected frequencies as a starting point.  This method incorporates the calculation of the covariance matrix which in turn gives us a measure of the precision of each parameter of the best-fit model.  This yielded more precise values of  $2.751190 \pm 0.000010$ mHz, $3.116709 \pm 0.000006$ mHz, $3.495113 \pm 0.000009$ mHz, and $4.443120 \pm 0.000012$ mHz, more than a 1000 fold increase in precision.

This new fitted sinusoidal function was then subtracted from the data and its periodogram can be found in the lower panel of Figure \ref{fig:cspec}.  The striking features of this de-signaled periodogram is the remainder of two signals of significant power near 4.44 mHz and 5.75 mHz.  However, the false-alarm probability arguments are not valid in a data set where signals have been removed artificially.  Interestingly, the excess power near 4.44 mHz is in a frequency bin significantly offset from our reported detected frequency.  If we treat both of these left over frequencies as real pulsation frequencies and use our algorithm, we find that there may exist two more detected frequencies at $4.450643 \pm 0.000017$ mHz and $5.755451 \pm 0.000018$ mHz.  The existence of these frequencies is questionable because neither frequency had enough power to reach our required false-alarm significance level in the full data set.  Additionally, the new 4.450643 mHz frequency is entirely lost within the window function around the original detected frequency.  When we de-signal the entire data set with the least-squares fitted sinusoidal function including the six frequencies, the resulting periodogram no longer contains any frequency bins with significant power.  The existence of these two pulsation frequencies is uncertain until better data with higher frequency sampling and more amicable window function can be obtained.

% -----------------------------------------------------------
% -----------------------------------------------------------

\section{Conclusions}
\label{sec:conc}

% FIGURE THREE

% FIGURE FOUR
\begin{figure*}
	\centering
	\epsscale{1.0}
	\plotone{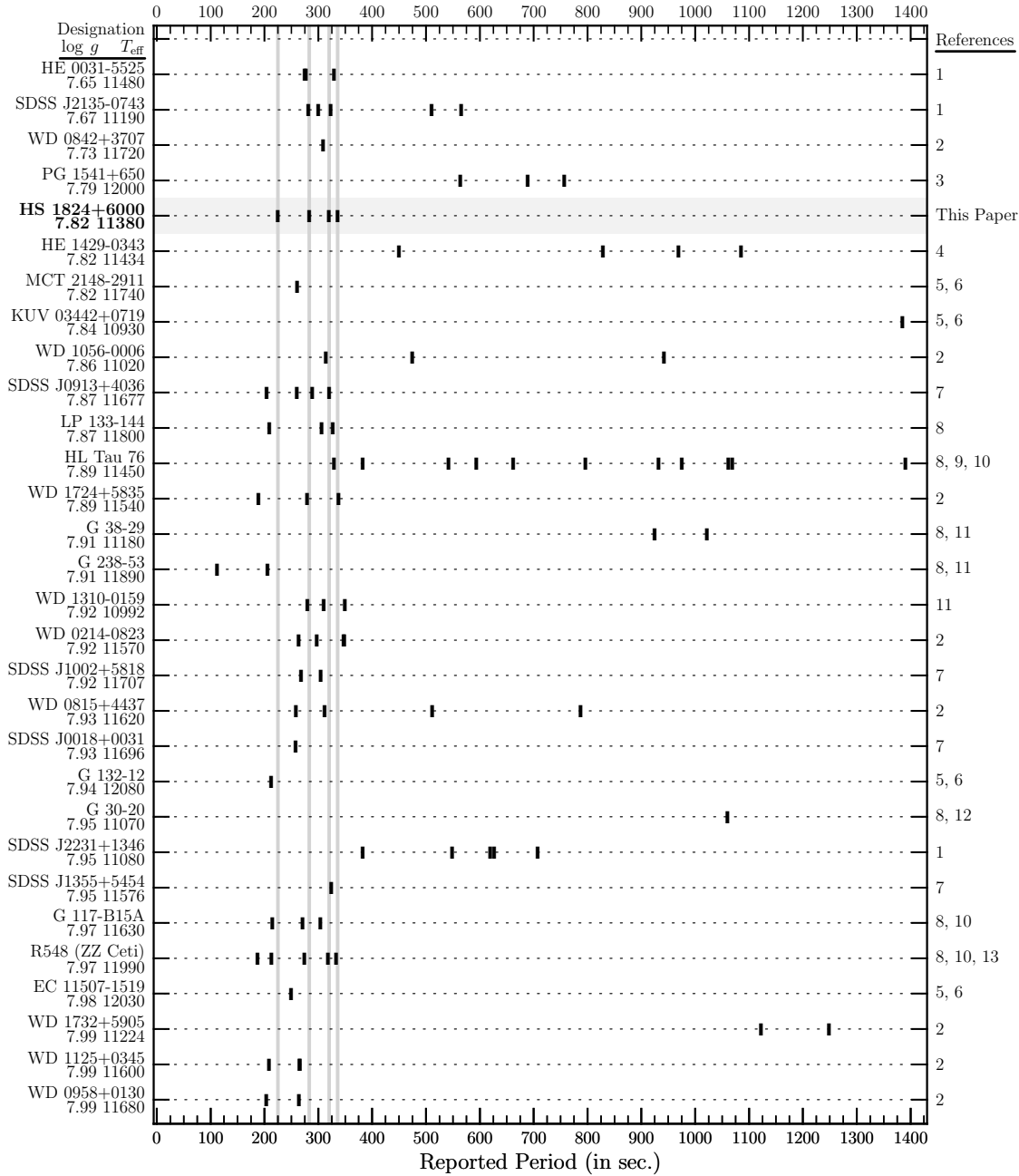}
	\caption{Spectrum of reported pulsation periods for published ZZ Ceti systems with spectroscopically determined $\log g < 8$.  HS 1824 is highlighted in gray with its four observed period locations marked with four vertical lines.  Some marks represent more than one (very closely spaced) observed pulsation period, see references for details.  HL Tau 76 lists only the verified independent pulsation modes of \citet{nd06}.  Several systems listed here are not included in Figure \ref{fig:instrip} as their $\log g$ and $T_{\rm eff}$ measurements are not of sufficient precision.
	References: 1  - \citealt{bgc06}, 2 - \citealt{asm04a}, 3 - \citealt{gv00}, 4 - \citealt{rs05}, 5 - \citealt{ag07}, 6 - \citealt{ag06}, 7 - \citealt{fm05}, 8 - \citealt{pb04}, 9 - \citealt{nd06}, 10 - \citealt{pb95}, 11 - \citealt{sok05b}, 12 - \citealt{asm02}, 13 - \citealt{asm03}.}
	\label{fig:perspec}
\end{figure*}

We have successfully detected four pulsation frequencies (periods), 2.751190 mHz (363.479 sec), 3.116709 mHz (320.851 sec), 3.495113 mHz (286.114 sec), and 4.443120 mHz (225.067 sec), in multiple observations of HS 1824+6000.  There are also two possible pulsation frequencies (periods) at 4.450643 mHz (224.687 sec) and 5.755451 mHz (173.748 sec).   With these periods of pulsation in HS 1824, the question remains if it, or other low gravity systems, can be empirically distinguished from the normal C/O core ZZ Ceti population.  To answer this we compiled all known ZZ Ceti stars with published pulsation periods and spectroscopically measured gravities of $\log g < 8.0$.  This search resulted in 30 systems including HS 1824.  In Figure \ref{fig:perspec} we plot all reported periods for these 30 systems.  Across all of these ZZ Ceti systems there exist many reported pulsation periods ranging from $100-1400$ sec.  However, it is apparent that better than half of the reported periods reside within the range of $150-400$ sec.  The four periods of HS 1824 are indistinguishable from the rest of this set of ZZ Ceti stars.  Further, there does not appear to be any distinction between the two low-mass ($\log g \lesssim 7.67$) systems (HE 0031-5525, SDSS J2135-0743, \citealt{bgc06}) and the rest of the set.  With this current set of data it appears that this empirical analysis of reported pulsation periods is not sufficient to distinguish a suspected He core from a normal C/O core.  However, HE 0031-5525, and SDSS J2135-0743 \citep{bgc06} are very close to the boundary of He and C/O core WDs and within the errors of their $\log g$ measurements may be C/O cores.

It remains  uncertain as to what degree this period spectrum comparative analysis can succeed. There are two primary differences between He and C/O core WDs that affect  g-modes: the contrast in mean molecular weights in their cores, and the one fewer stratified layer in a He core object.  G-modes penetrate deeply into the core, so that  differences in the Brunt-V\"{a}is\"{a}l\"{a} profile there (due to the mean molecular weight; see \citealt{cjd02}) significantly change the resulting mode period spectrum \citep{pa06}. The stratified layers of material within the WD also affects how different pulsation modes are trapped, driven, and excited \citep{ahc02,pa06}. He core WDs possess only two zones of He and H, while C/O core WDs possess the additional zone of C/O. Qualitatively, both of these differences would produce differences in the mode period spectra, and are the subject of current theoretical work we are pursuing. Once  these full mode calculations are available, we can answer whether clear differences are observable.

Most reported systems in Figure \ref{fig:perspec} were found in observational campaigns looking only for pulsations in an effort to constrain the ZZ Ceti instability strip.  In most cases, no attempt was made to distinguish observed pulsation periods as independent modes, as opposed to linear combinations of modes.  This analysis was neglected in large part due to the lack of extensive follow up.  Our observations of HS 1824 showed most single nights of data contain the pulsations of one specific period and it was a rarity to find a night of data with multiple pulsation periods.  Ideally, very long gapless observations on the order of several days would address these problems very well.  These observations could be obtained through the use of telescope networks such as the Whole Earth Telescope\footnote{http://www.physics.udel.edu/darc/wet} as was done with HL Tau 76 \citep{nd06} and G117-B15A \citep{sok91,sok95} and the Las Cumbres Observatory Global Telescope\footnote{http://www.lcogt.net}.  We look toward future, more detailed observations of many low-mass and normal-mass ZZ Cetis to help provide a measurable distinction between He and C/O core compositions in WDs.

\acknowledgments

We thank Anjum Mukadam for alerting us to the existence of this object and the referee for comments that clarified our presentation.  We thank Phil Arras for useful discussion on pulsations in He core WDs.  This work was supported by the National Science Foundation under grants PHY 05-51164 and AST 07-07633.

% -----------------------------------------------------------
% -----------------------------------------------------------

\end{document}